# Formation of Co nanoclusters in epitaxial $Ti_{0.96}Co_{0.04}O_2$ thin films and their ferromagnetism

D. H. Kim, J. S. Yang, K. W. Lee, S. D. Bu, and T. W. Noh[a]

*ReCOE and School of Physics, Seoul National University, Seoul 151-747, Korea*

S.-J. Oh

*CSCMR and School of Physics, Seoul National University, Seoul 151-747, Korea*

Y.-W. Kim

*School of Materials Science and Engineering, Seoul National University, Seoul 151-744, Korea*

J.-S. Chung

*Department of Physics, Soongsil University, Seoul 156-743, Korea*

H. Tanaka, H. Y. Lee, and T. Kawai

*ISIR, Osaka University, Osaka 567-0047, Japan*

Anatase $Ti_{0.96}Co_{0.04}O_2$ films were grown epitaxially on $SrTiO_3$ (001) substrates by using plused laser deposition with *in-situ* reflection high-energy electron diffraction. The oxygen partial pressure, $P_{O2}$, during the growth was systematically varied. As $P_{O2}$ decreased, the growth behavior was changed from a 2-dimensional layer-by-layer-like growth to a 3-dimensional island-like one, which resulted in an increase in the saturation magnetization. These structural and magnetic changes were explained in terms of the formation of cobalt clusters whose existence was proved by transmission-electron-microscopie studies. Our work clearly indicates that the cobalt clustering will cause room-temperature ferromagnetism in the Co-doped $TiO_2$ films.

---

[a] Corresponding Author; e-mail: twnoh@phya.snu.ac.kr

There has been lots of attention paid to dilute magnetic semiconductors (DMS) due to their application potential in the rapidly developing area of spintronics.[1-3]. The fundamental issues, such as the origin and the nature of ferromagnetism (FM), are also intriguing enough to attract attention from many scientists. While carrier-induced interaction between the magnetic ions has been suggested as the possible origin for the FM in DMS, the precise mechanism is still controversial.[3]

Recently, Matsumoto et al.[4] reported occurrence of room temperature FM in Co-doped anatase $TiO_2$ ($Co:TiO_2$) films which were grown by the combinatorial laser molecular beam epitaxy (MBE) techniques. This work generated great interest in the DMS community, since it could open possibilities to create new multifunctional oxides which utilize the spin polarized carriers.[5-10] Chambers et al.[6] reported the magnetic and structural properties of the $Co:TiO_2$ films grown by the oxygen plasma assisted MBE and claimed that the FM should be intrinsic. However, Shinde et al.[8] measured the ion-channeling spectra and temperature-dependent magnetization of the $Co:TiO_2$ films grown by the pulsed laser deposition, but suggested that the FM most likely arises from Co nanocluster formation.

In this letter, we describe our systematic investigation of the magnetic and microstructural properties of the $Co:TiO_2$ films grown under various oxygen partial pressures ($P_{O2}$). At $P_{O2}$ values higher than $1.0 \times 10^{-5}$ torr, we could grow high quality epitaxial films with small magnetization values. As $P_{O2}$ values decreased, the conductivity and the magnetization of the films increased systematically. Using numerous characterization methods, including cross-sectional transmission electron microscopy (XTEM) and energy-dispersive spectroscopy (EDS), we showed that the observed $P_{O2}$ dependences in our films could be explained by the Co nanoclusters embedded in the epitaxial $TiO_2$ films.

To grow high quality epitaxial $Ti_{0.96}Co_{0.04}O_2$ films on $SrTiO_3$ (001) single crystalline substrates, we used a pulsed laser deposition (PLD) method with *in-situ* reflection high energy electron diffraction (RHEED), which is known as "laser MBE".[11] A sintered poly-crystalline $Ti_{0.96}Co_{0.04}O_2$ target was ablated by a KrF excimer laser (wavelength: 248 nm) with a fluence of 1.5 J/cm$^2$ at 2 Hz. The substrate temperature was maintained at 600 °C and $P_{O2}$ was varied from $1.0 \times 10^{-7}$ to $1.0 \times 10^{-4}$ torr. The growth rate was controlled to be extremely slow, less than 0.2 nm/min, which was reported to be necessary to reduce the Co inhomogeneity.[6,7] The thickness of each film was calculated from the oscillations of x-ray reflectivity curves and found to be about 70 nm.

The growth behavior of the $Ti_{0.96}Co_{0.04}O_2$ films was found to be strongly dependent on $P_{O2}$. As shown in Fig. 1(a), the films grown at $P_{O2} \geq 1.0 \times 10^{-5}$ torr showed clear streaky RHEED patterns, which suggest 2-dimensional (2D) layer-by-layer growth with very smooth surfaces. *Ex-situ* atomic force microscope (AFM) studies showed that their root-mean-square roughness should be a few nanometers. For the films grown at lower $P_{O2}$, their RHEED patterns were similar to Fig. 1(a). However, as the film growth progressed, the patterns turned into 3D spotty patterns, as shown in Fig. 1(b). The 3D island growth was also confirmed by AFM, and the width and height of the islands were found to vary from 10 to 100 nm.

Figure 1(c) shows a typical x-ray diffraction (XRD) spectrum for the $Ti_{0.96}Co_{0.04}O_2$ films. The $2\theta$-$\theta$ scan shows only (00l) peaks of the anatase phase, while other peaks from different phases or orientations cannot be seen. The rocking-curve of the (004) peak for the film grown at $1.0 \times 10^{-5}$ torr shows a full width at half maximum (FWHM) of 0.66°, which value is similar to those reported by Murakami *et al.*[12] As $P_{O2}$ decreased down to $1.0 \times 10^{-7}$ torr, the FWHM increased to 0.86°, indicating that the film grown under the low $P_{O2}$ show a wider mosaic spread. This change can be related to the inhomogeneity of Co in the films grown at the low $P_{O2}$, as will be shown later.

The magnetic and the transport properties of the anatase $Co:TiO_2$ films showed systematic trends in relation to $P_{O2}$. The magnetization of the films was measured with a superconducting quantum interference device. As shown in Fig. 2(a), most of the films, except that grown at $1.0 \times 10^{-4}$ torr, showed FM at room temperature. As $P_{O2}$ decreases, the saturation magnetization increases rather systematically, as shown in Fig. 2(b). Similarly, the conductivity of the films also increases as $P_{O2}$ decreases.[13] These systematic changes might be interpreted in terms of FM induced by carriers between the magnetic Co ions.

However, the inhomogeneous distribution of the Co ions should be carefully checked for films with a meta-stable phase, as in our anatase Co:TiO$_2$ films. Note that TiO$_2$ can have three different phases (i.e., rutile, anatase, and brookite) and that the most stable phase is the rutile. Although the anatase phase is difficult to make in bulk, it is meta-stable enough to be grown in thin film form on lattice matching substrates.[12] Because the anatase is not thermodynamically stable, defects can be easily formed with a slight change in the film growth conditions. In addition, the Co ions are known to diffuse easily at a relatively low temperature, such as 400°C.[14] Therefore, the possibility of Co and/or Co-intermetallic clusters should be considered carefully in the case of oxide films grown at higher temperatures.

We made some experimental observations which should be interpreted in terms of the formation of the Co clusters. As shown in Fig. 2(b), most of the films deposited at $P_{O2}$ under $3.0 \times 10^{-5}$ torr had saturation magnetization values close to that of bulk cobalt (1.7 $\mu_B$). Also, the magnetization vs. temperature data, shown in the inset of Fig. 2(b), shows that FM persists up to 750 K. Especially, for the film grown at $P_{O2} = 1.0 \times 10^{-7}$ torr, the magnetization does not decrease much with temperature. Note that the Curie temperature of metal cobalt is 1404 K. This high lower bound of the Curie temperature is rather difficult to explain in terms of the carrier-induced FM. We also performed x-ray magnetic circular dichroism (XMCD) and magneto-optical Kerr effect measurements on several Co:TiO$_2$ films, and all of the ferromagnetic samples had XMCD spectra close to that of bulk cobalt.[14]

In order to get more direct evidence of the formation of the Co clusters, we performed XTEM studies. As shown in Fig. 3(a), the high resolution XTEM images of the films grown at $P_{O2}$ above $1.0 \times 10^{-5}$ torr showed high quality epitaxial films without any sign of Co inhomogeneity. On the other hand, the XTEM picture of the film grown at $3.3 \times 10^{-7}$ torr, shown in Fig. 3(b), revealed some nanoclusters inside the film. The clusters at the interface were larger in diameter than the ones inside the film with an overall average diameter of 11.7 ± 6.0 nm. The apparent horizontal dark line, in the middle of the Fig. 3(b), comes from the strain field induced by the lattice mismatch between the film and the substrate. An EDS mapping of elemental Co, in Fig. 3(c), clearly indicates that the nanoclusters are mainly composed of Co. The density of the Co clusters is much higher at the interface, which can work as preferential nucleation sites due to its instability. The high resolution XTEM image near the interface area is shown in Fig. 3(d). At the top of the circular particle, Moiré fringes are visible, suggesting that two different crystals were overlapping. Under the higher magnification, the circular cluster has lattice fringes with a spacing of 1.92 Å, which is equivalent to the $(10\bar{1}1)$ plane of metal Co.

The dependence of the magnetic properties of the Co:$TiO_2$ films, shown in Fig. 2, on $P_{O2}$ can be easily understood in terms of the Co nanoclusters. Note that, from the thermodynamic point of view, cobalt oxide is less stable than $TiO_2$ at a low $P_{O2}$. The oxygen vacancies in the anatase $TiO_2$ films grown at low $P_{O2}$ will help the diffusion of the Co ions, resulting in the formation of the nanoclusters. As the number of Co clusters increases, the saturation magnetization will become larger. Although our TEM and XMCD work suggests that the nanoclusters should be mainly composed of Co, the cluster formation of other Co intermetallic compounds is not completely ruled out. Note that, as shown in Fig. 2(b), the value of saturation magnetization for the film grown at $P_{O2} = 1.0 \times 10^{-7}$ torr is somewhat larger than that of bulk Co metal, which might be related to the formation of Co intermatallic compounds. Further studies are required to clarify this possibility.

In summary, we grew high quality anatase $Ti_{0.96}Co_{0.04}O_2$ films at various oxygen partial pressures. As the pressure decreased, the growth behavior changed from a 2-dimensional layer-by-layer like growth to a 3-dimensional island one, resulting in a rough surface and wider mosaic spread of the film. Also, the saturation magnetization increased. These changes in microstructural and magnetic properties could be explained by the formation of cobalt nanoclusters, whose existence was identified by cross-sectional TEM. Although the possibility of other intrinsic origins cannot be completely ruled out, this work strongly suggests that the formation of Co clusters should be the main origin for the observed ferromagnetism at least in our films.

We wish to thank J. H. Park, Y. D. Park, and J.-G. Yoon for valuable discussions. We also thank H. C. Kim and H. C. Ri for the initial magnetization measurements at KBSI. This work was financially supported by KOSEF through CSCMR, and by the Ministry of Science and Technology through the Creative Research Initiative program.

Figure Captions:

Fig. 1 The *in-situ* RHEED patterns of $Ti_{0.96}Co_{0.04}O_2$ films which were grown at $P_{O2}$ of (a) $1.0 \times 10^{-5}$ torr and (b) For $1.0 \times 10^{-7}$ torr. (c) The XRD pattern for a $Co:TiO_2$ film on $SrTiO_3$ (001).

Fig. 2 (a) Magnetization *vs.* magnetic field curves at 300 K for the $Ti_{0.96}Co_{0.04}O_2$ thin films under various partial oxygen pressure, $P_{O2}$. The data for the films grown at $1.0 \times 10^{-7}$, $3.3 \times 10^{-7}$, $1.0 \times 10^{-6}$, $3.3 \times 10^{-6}$, $1.0 \times 10^{-5}$, and $1.0 \times 10^{-4}$ torr are shown as the solid, dashed, dash-dotted, dotted, and short dotted curves, respectively. (b) Dependence of the saturation magnetization, *Ms*, on $P_{O2}$. The dotted line corresponds to the bulk cobalt metal value. The inset shows temperature dependence of the normalized magnetization for the films grown at a $P_{O2}$ of $1.0 \times 10^{-7}$ torr (the squares) and $1.0 \times 10^{-5}$ torr (the triangles).

Fig. 3 (a) A high resolution XTEM picture of the $Ti_{0.96}Co_{0.04}O_2$ film grown at a $P_{O2}$ of $1.0 \times 10^{-5}$ torr. It can be seen that the high quality epitaxial film was grown on a $SrTiO_3$ substrate. (b) A low resolution XTEM picture and (c) the corresponding EDS mapping of the Co atom for the film grown at $1.0 \times 10^{-7}$ torr. It can be easily seen that Co-rich nanoclusters were formed. (d) A high resolution XTEM picture of the nanoclustered film near the film/substrate interface.

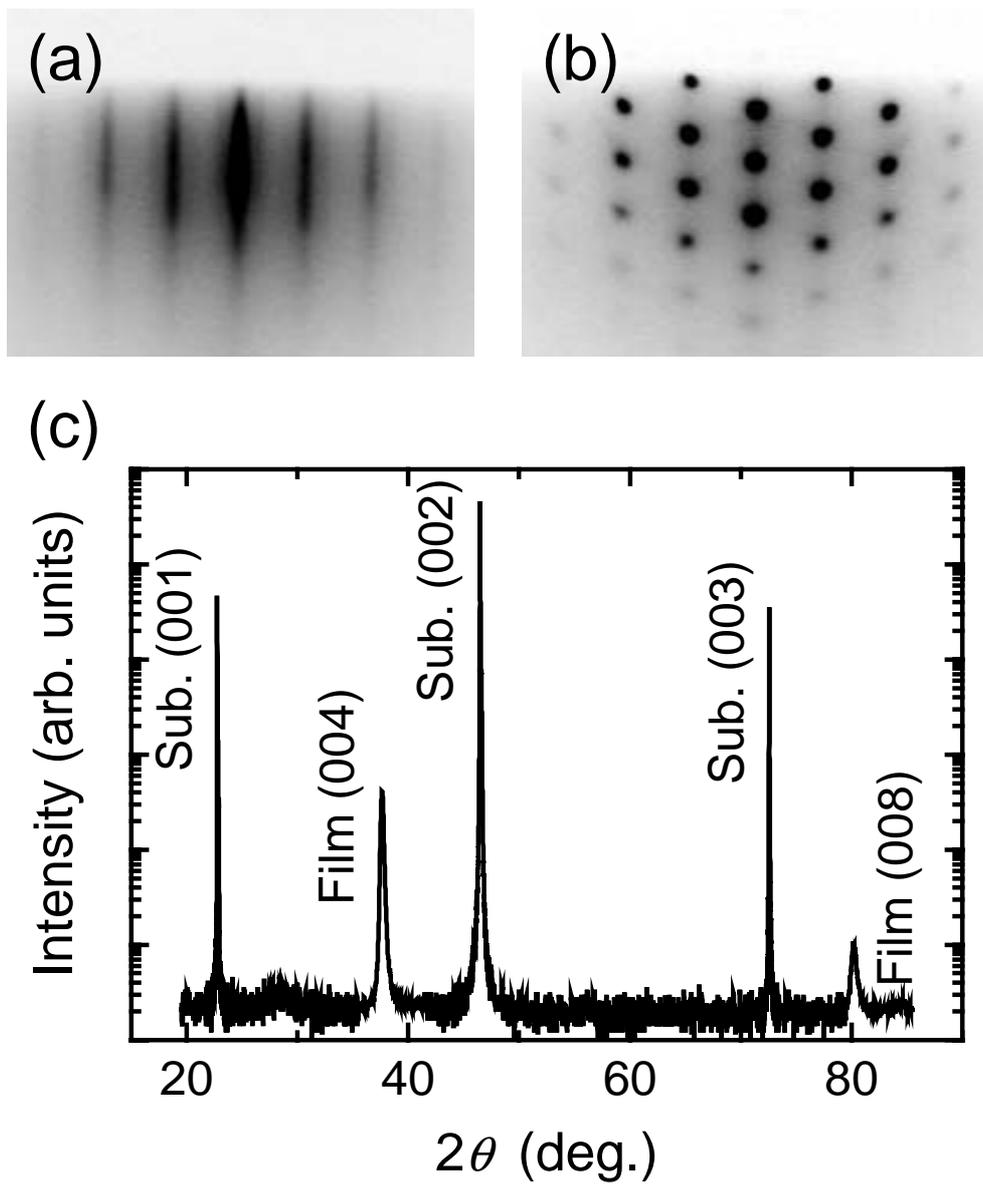

Fig. 1 D.H. Kim *et al.*

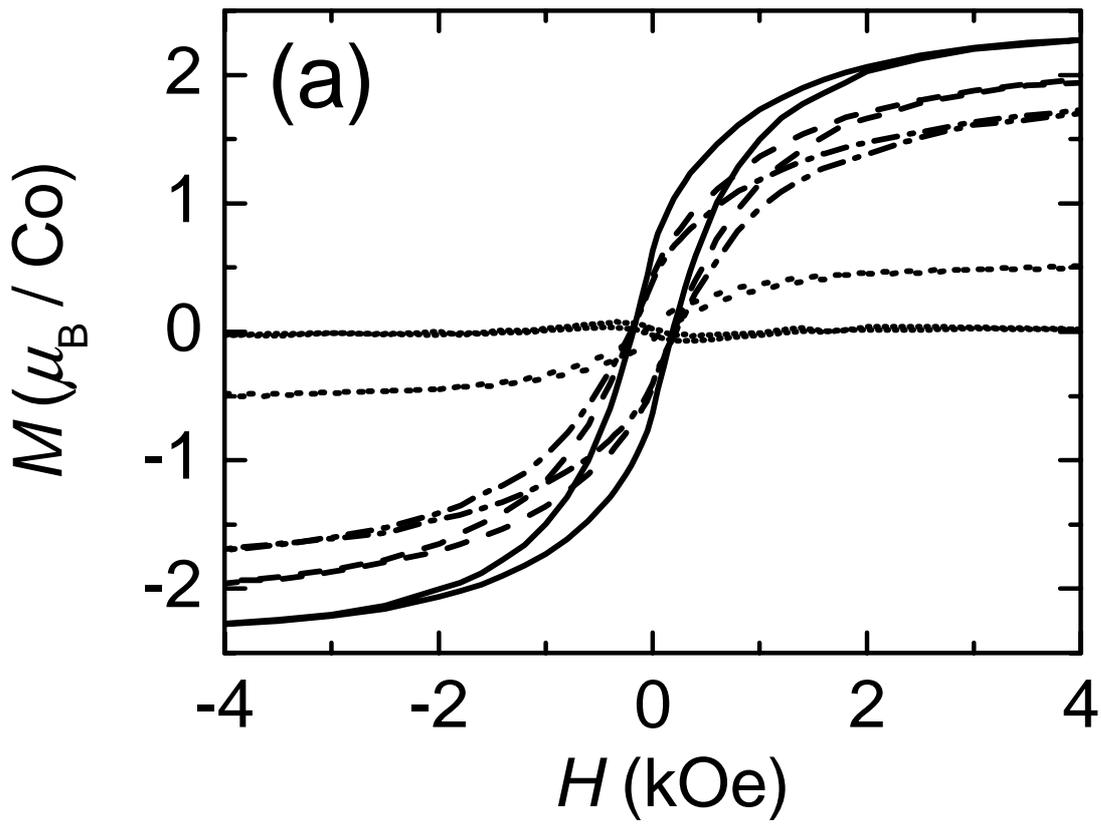
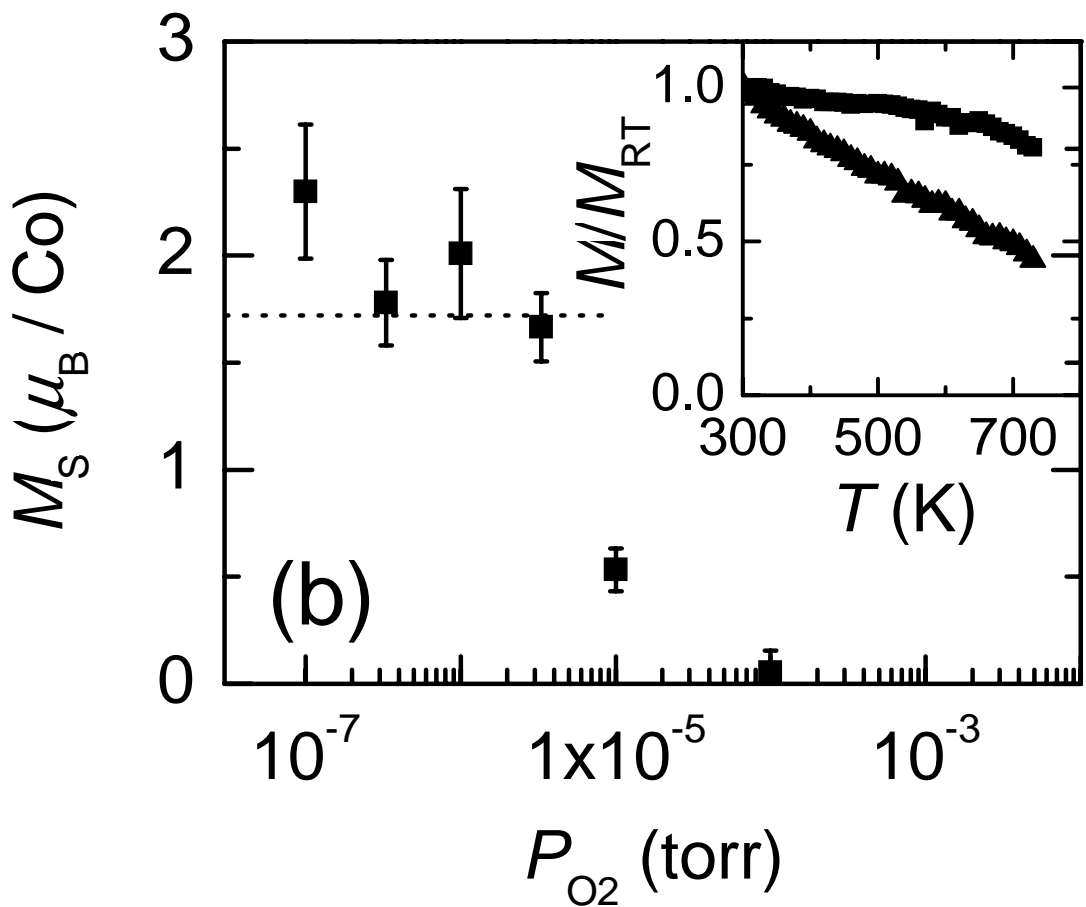

Fig. 2  D.H. Kim *et al.*

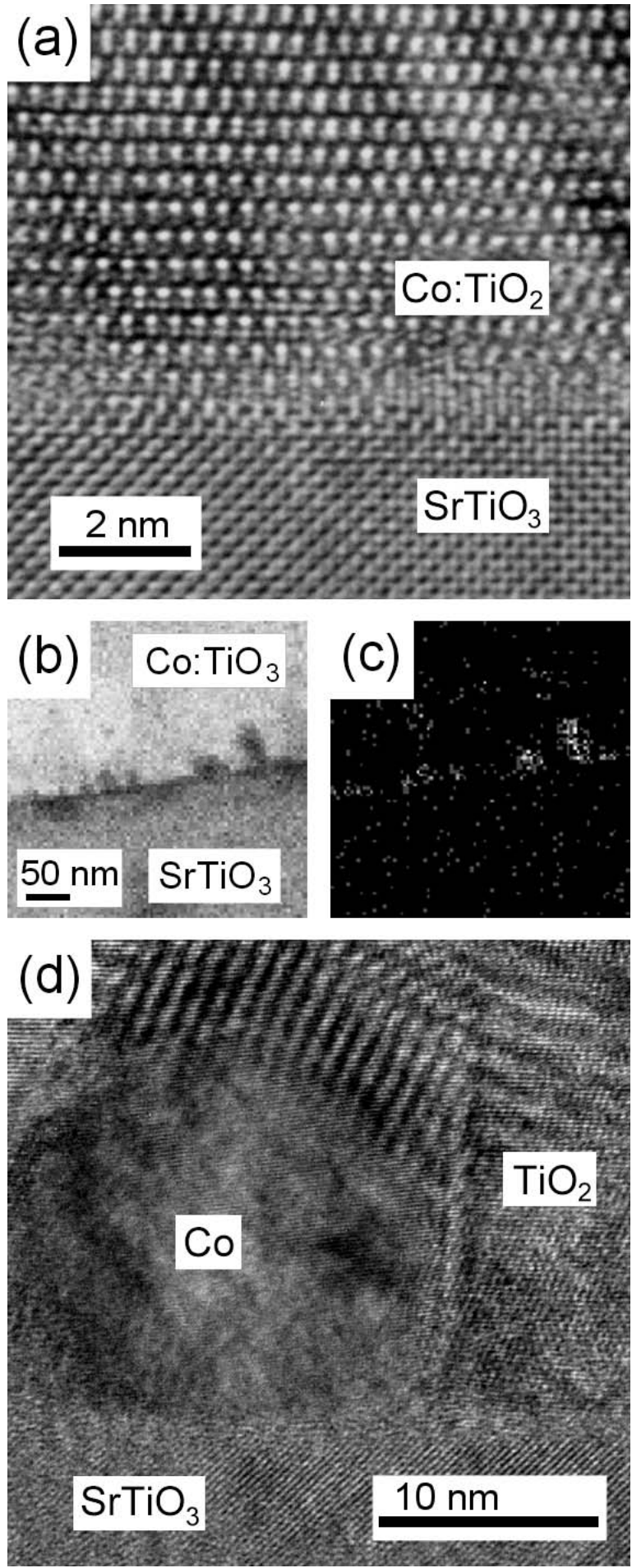

Fig. 3     D. H. Kim *et al.*